# High-current *p*-type transistors from precursor-engineered synthetic monolayer WSe$_2$

Anh Tuan Hoang[1,†], Kathryn Neilson[2,†], Kaikui Xu[3], Yucheng Yang[3], Stephanie M. Ribet[4], Tara Peña[2], Giulio D'Acunto[5], Young Suh Song[2], Anton E. O. Persson[2], William Millsaps[1], Colin Ophus[1,4], Matthew R. Rosenberger[3], Eric Pop[1,2,7*], Andrew J. Mannix[1,6*]

[1] *Department of Materials Science & Engineering, Stanford University, Stanford, CA 94305, USA*
[2] *Department of Electrical Engineering, Stanford University, Stanford, CA 94305, USA*
[3] *Department of Aerospace and Mechanical Engineering, University of Notre Dame, Notre Dame, Indiana 46556, USA*
[4] *National Center for Electron Microscopy (NCEM), The Molecular Foundry, Lawrence Berkeley National Laboratory, Berkeley, CA 94720, USA*
[5] *Department of Chemical Engineering, Stanford University, Stanford, CA 94305, USA*
[6] *Stanford Institute for Materials & Energy Sciences, SLAC National Accelerator Lab., Menlo Park, CA 94025, USA*
[7] *Department of Applied Physics, Stanford University, Stanford, CA 94305, USA*

[†] *These authors contributed equally.*

*Contact:* epop@stanford.edu, ajmannix@stanford.edu

**Abstract:**

Monolayer tungsten diselenide (WSe$_2$) is a leading candidate for nanoscale complementary logic. However, high defect densities introduced during thin-film growth and device fabrication have limited *p*-type transistor performance. Here, we report a combined strategy of precursor-engineered chemical vapor deposition and damage-minimizing fabrication to overcome this limitation. By converting tungsten trioxide and residual oxyselenides into reactive suboxides before growth, and precisely regulating selenium delivery during deposition, we synthesize uniform, centimeter-scale monolayer WSe$_2$ films with charged defect densities as low as $5 \times 10^9$ cm$^{-2}$. Transistors fabricated from these films achieve record *p*-type on-state current up to 888 µA·µm$^{-1}$ at $V_{DS}$ = -1 V, matching leading *n*-type devices. This leap in material quality closes the *p*-type performance gap without exotic doping or contact materials, marking a critical step towards complementary two-dimensional semiconductor circuits.

# Introduction

For over half a century, aggressive transistor scaling and improved electrostatic control have driven advances in silicon (Si) complementary metal oxide semiconductor (CMOS) logic. Future improvements in Si CMOS face fundamental challenges because additional thinning of the Si channel degrades carrier mobility through increased surface scattering.[1] Monolayer two-dimensional (2D) semiconductors such as the transition metal dichalcogenides (TMDs, e.g. $MoS_2$ or $WSe_2$) offer an atomically thin channel with superior electrostatic control and reduced surface scattering in sub-nanometer films.[2] They can also be synthesized by back-end-of-line-compatible processes (below ~400 °C) on arbitrary substrates,[3] enabling high-density, stacked heterogeneous integration.[4]

Unlike other next-generation materials such as oxide semiconductors, TMDs inherently offer both *n*-type and *p*-type transport, which is essential for low-power CMOS.[2] However, their transistor performance remains below theoretical limits due to defects and impurities present in the available materials.[5] Monolayers with reduced defect densities are typically obtained by exfoliation from bulk crystals,[6] which limits scalability. Considerable effort has therefore focused on large-area growth,[7] with chemical vapor deposition (CVD) being a promising, industry-compatible route.[8] However, defect control in CVD remains challenging, and the role of precursor chemistry is underexplored.[9] Approaches such as tungsten foils, alternative chalcogen sources, or growth promoters can reduce defect densities at the expense of potential contamination.[10-12] Selenide TMDs are particularly difficult: they are more prone to oxidation and Se-vacancy formation,[13,14] and Se precursors typically have lower vapor pressure, requiring higher vaporization temperatures that complicate flow control and degrade uniformity.

Here, we report a CVD strategy for monolayer $WSe_2$ that addresses these longstanding challenges, producing films with exceptional uniformity and crystalline quality that exhibit record *p*-type transistor performance. By thermodynamically and kinetically optimizing the growth environment – including pre-conversion of tungsten oxide to suboxides, removing oxyselenides, and precisely controlling Se partial pressure – we synthesize monolayer $WSe_2$ with a record-low charged defect density of $(5.2 \pm 1.1) \times 10^9$ cm$^{-2}$. This is comparable to the best exfoliated materials today.[15] Using an optimized, damage-minimizing fabrication flow, we realize monolayer $WSe_2$ transistors with record-high *p*-type on-current densities up to 888 µA µm$^{-1}$ at $V_{DS}$ = -1 V, comparable to state-of-the-art *n*-type $MoS_2$ transistors.[16] Notably, this performance is achieved without advanced doping or contact materials, underscoring the quality of the $WSe_2$ channel.

## Synthesis of uniform monolayer $WSe_2$

**Figure 1a** illustrates our solid-source CVD (SSCVD) method for synthesizing uniform monolayer $WSe_2$ from $WO_{3-x}$ and elemental Se in a two-zone tube-furnace system, incorporating a pre-processing step to optimize the chemical state of the tungsten precursor. Growth proceeds by placing a clean sapphire substrate on the platen, performing pump-purge cycles to remove residual air, then heating the tube under argon carrier gas flow. Selenium vapor is delivered by thermal dissociation of Se powder in upstream Zone 1, while heating Zone 2 provides the tungsten precursor flux and thermal energy for deposition.

SSCVD growth typically requires reactor seasoning,[17] in which the fused silica sample platen and reactor walls are passivated by a thin layer of tungsten oxyselenide ($WO_xSe_y$). Film quality is critically sensitive to this step. While SSCVD commonly supplies the tungsten precursor by evaporating powder from a crucible boat, we obtain superior results by evenly spreading the precursor as a thin layer on the platen and annealing the tube and platen at 1000 ºC under argon flow. This platen recovery cycle (PRC) is repeated after every growth, with the chemical evolution of the seasoning layer analyzed in a later section. Growth without the PRC produces isolated $WSe_2$ crystals with multilayer nucleation and variable domain sizes (**Figure 1b** and **Supplementary Figure 1**). In contrast, applying the PRC yields large-area, continuous monolayer $WSe_2$ (**Figure 1c,d**), with atomic force microscopy (AFM) confirming uniform thickness and a clean, particle-free surface (**Figure 1e**).

To assess film uniformity, Raman spectra were measured over a $1 \times 1$ cm$^2$ area of monolayer $WSe_2$ on sapphire, grown with the PRC (**Figure 1f**). The absence of the $B_{2g}$ peak at ~310 cm$^{-1}$ confirms monolayer thickness, while the E'+A$_1$' mode shows a narrow distribution centered at $251 \pm 0.5$ cm$^{-1}$ with minimal spatial variation (**Figure 1g,h**), indicating uniform thickness and strain across the centimeter-scale film. Photoluminescence (PL) mapping over a $50 \times 50$ μm$^2$ region of this sample revealed sharp neutral exciton emission (**Figure 1i**) with a full-width at half-maximum (FWHM) of $39 \pm 7$ meV, showing a narrow histogram (**Figure 1j**) and uniform spatial distribution (**Figure 1k**). These values are consistent with high crystalline quality, matching uncapped, exfoliated monolayers on sapphire at room temperature.[18]

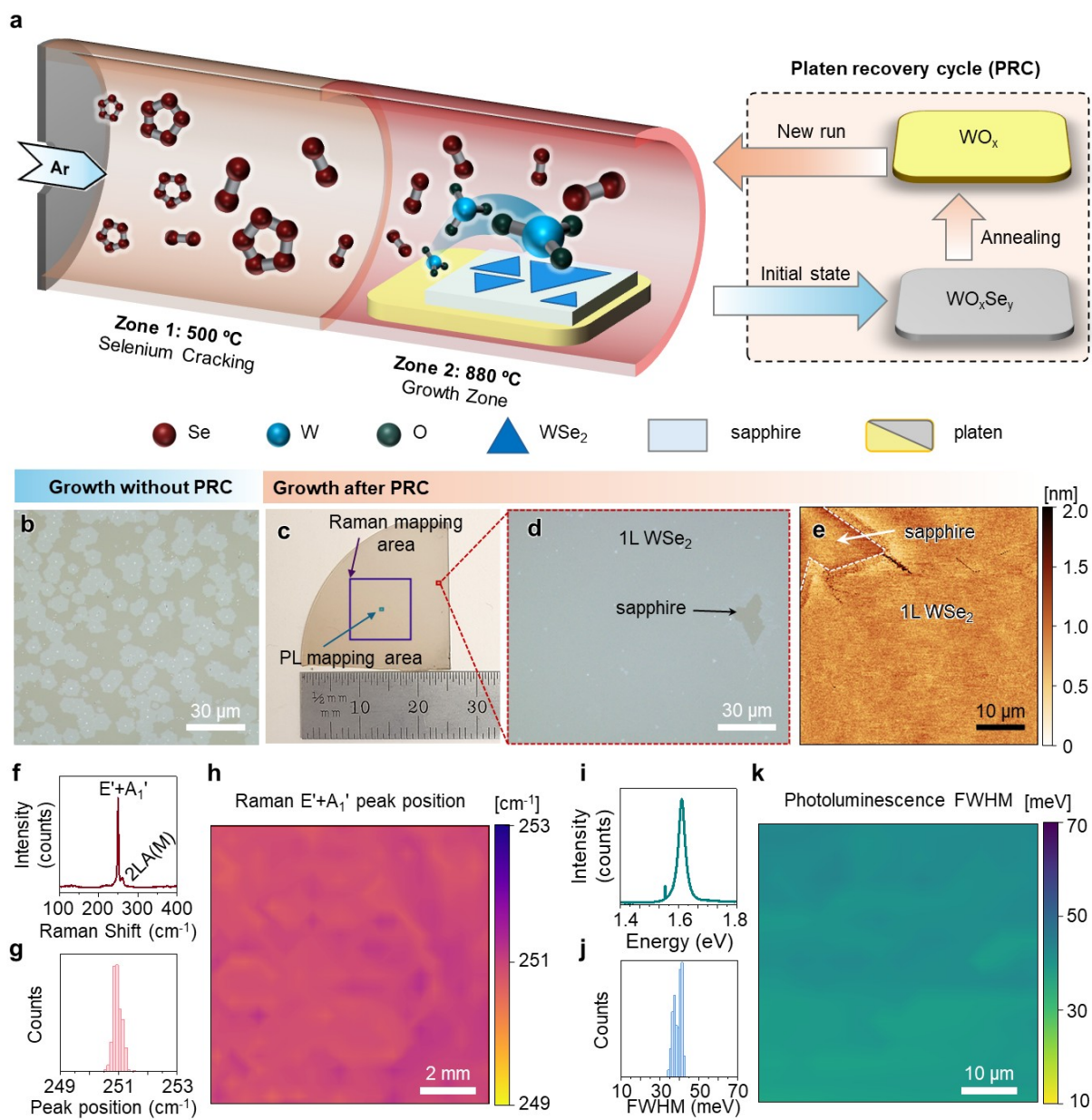

**Figure 1: Synthesis and characterization of uniform monolayer WSe$_2$. a,** Schematic of WSe$_2$ synthesis in two-zone furnace with the platen recovery cycle (PRC). **b,** Optical microscope image of isolated WSe$_2$ domains obtained from synthesis without the PRC. **c,** Photograph of uniform large-area monolayer (1L) WSe$_2$ on sapphire, synthesized after the PRC. **d,** Optical microscope image from an area near the wafer edge, with a small gap revealing the substrate. **e,** Atomic force microscopy (AFM) topography image, showing the smooth, particle-free surface. **f,** Representative Raman spectrum of as-grown WSe$_2$ on sapphire. **g,** Histogram of E'+A$_1$' peak position from a 1 × 1 cm$^2$ map (25 × 25 array point measurement), with a narrow distribution at 251 ± 0.5 cm$^{-1}$. **h,** Colormap of E'+A$_1$' peak position over the same area. **i,** Typical photoluminescence (PL) spectrum of monolayer CVD WSe$_2$. **j,** Histogram of PL full-width at half-maximum (FWHM) from a 50 × 50 µm$^2$ map (15 × 15 points), showing a narrow FWHM of 39 ± 7 meV. **k,** Colormap of PL FWHM over the same area, confirming the microscale uniformity and monolayer thickness of the WSe$_2$ film.

## Elucidating the platen recovery mechanism for improved WSe$_2$ growth

**Figure 2** shows how changes in precursor composition and surface chemistry during the PRC correlate with improved WSe$_2$ film quality. Photographs compare the as-purchased WO$_3$ powder precursor (**Figure 2a**) with the platen surface at successive PRC stages (**Figures 2b-d**). The shiny grey surface observed immediately after a growth run (**Figure 2b**) differs markedly from the WO$_3$ powder. Although this transformed reactor surface is rarely examined in the literature, it plays a critical role in determining the resulting WSe$_2$ film quality. Annealing the platen and furnace tube at 1000 °C under argon flow modifies the seasoning layer, yielding the intermediate-stage (short PRC anneal, **Figure 2c**) and final-stage surfaces (optimized PRC anneal, **Figure 2d**). Only growths using the platen at this final stage produced complete monolayer coverage (**Supplementary Note 1** and **Supplementary Figure 1**).

To understand the underlying mechanism, we measured the chemistry at each platen stage using X-ray photoelectron spectroscopy (XPS). The W 4f core-level spectra were fitted with doublets corresponding to WO$_3$ (37.1 eV and 34.9 eV), and WSe$_2$ (34.8 eV and 31.8 eV) (**Figure 2e** and **Supplementary Figure 2**). Compared to reference WO$_3$ powder, the initial-stage platen shows a strong W-Se component (78 % of the total W 4f intensity), consistent with extensive WSe$_2$ formation and residual WO$_{3-x}$. At the intermediate stage, the W-Se component drops to ~10%, accompanied by the emergence of a W-O loss feature at 40.8 eV, characteristic of sub-stoichiometric WO$_{3-x}$.[19] In the final stage, no detectable W-Se signal remains. Only the WO$_{3-x}$ doublet and the W-O loss feature are observed. This progression suggests that selenium removal from the platen during the PRC is essential for achieving uniform monolayer growth. The Se 3d core-level spectra, shown in **Figure 2f** (at 54.3 eV and 53.5 eV), confirm the same trend.

The precise chemical state of the WO$_x$ precursor also matters. Comparing W 4f spectra of the as-received powder with the final-stage platen surface shows little difference in line shape (**Supplementary Figure 2**). However, the O 1s spectrum reveals a broader peak for the final-stage surface (**Figure 2g**), which we attribute to oxygen vacancies that shift the O 1s core-level photoemission toward higher binding energies.[20] This interpretation is supported by the broadened photoluminescence of PRC annealed WO$_{3-x}$ compared to the starting WO$_3$ powder (**Supplementary Figure 3**).[20]

Together, these results suggest the following mechanism. During SSCVD growth, elemental Se (mainly Se$_2$, **Supplementary Figure 4**) reacts with WO$_3$ to generate an amorphous tungsten-oxyselenide (WO$_x$Se$_y$) layer which coats the platen and furnace tube. Mixed Se–O ligands in this layer drastically increase tungsten volatility.[21] Above ~850 °C, any WO$_x$Se$_y$ present during growth decomposes into W suboxides and vapors of Se$_x$O$_y$ and Se$_2$. Introducing this Se$_x$O$_y$ to the growing WSe$_2$ film surface depletes the reactive free-Se reservoir and injects reactive oxide species into the boundary layer – the diffusion space above the substrate during thin film growth.[22] This Se-poor, O-rich environment lowers the formation energy of positively charged Se vacancies ($V_{Se}^+$) and promotes exothermic O substitution on chalcogen sites ($O_{Se}$), which act as abundant acceptors in TMDs.[23] Furthermore, amorphous $WO_xSe_y$ has been reported to suppress 2H $WSe_2$ nucleation,[24] consistent with the poor reproducibility observed from growth without the PRC.

The PRC protocol, annealing the reactor and platen at 1000 °C, removes these deleterious $Se_xO_y$ species and converts the coating to oxygen-deficient $WO_{3-x}$. This treatment enables a near-stoichiometric Se:W flux and improves nucleation conditions. Sub-stoichiometric $WO_{3-x}$ also acts as a more favorable precursor for $WSe_2$ growth[25] due to its lower oxidation state and enhanced reactivity, producing films with lower defect density and greater uniformity.[26,27] Evenly distributing $WO_{3-x}$ across the platen further promotes uniform precursor delivery (**Figure 1c-d** and **Supplementary Figure 5**).

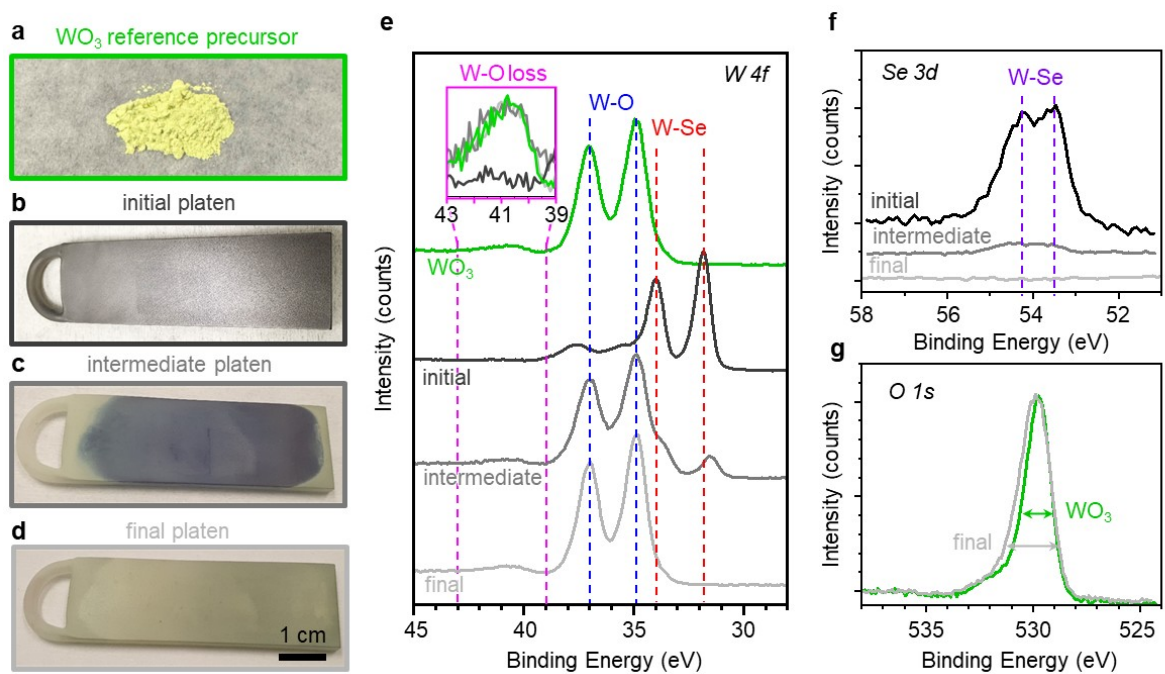

**Figure 2: Preparation and oxidation state of the platen leading to high-quality growth. a**, Optical image of $WO_3$ reference powder, as received, prior to synthesis. **b**, The initial state of the platen prior to synthesis of a new chip, but following a prior growth, containing $WO_xSe_y$ coating layer. **c**, The intermediate stage of the platen, after an incomplete anneal at 1000 °C, where some (but not all) of the $WO_xSe_y$ still remains. **d**, The final stage of the platen, after annealing at 1000 °C has completed. At this stage, the platen is ready for high-quality $WSe_2$ synthesis. The fused silica platen measures 10 cm in length, 3 cm in width. **e**, X-ray photoelectron spectroscopy (XPS) scan of W 4f core level, for **a-d** (platen scans are taken on a smaller, reference alumina substrate to fit into the XPS chamber). Inset shows the magnified views of the W-O loss feature. **f,** Se 3d core level spectra. Significant amounts of Se are present prior to annealing, and only after completion of the 1000 °C anneal does the Se signal become unresolvable. **g,** O 1s core level spectra, where the broadening of the emission suggests additional oxygen vacancies.

## Improved monolayer *p*-type transistors

We evaluated the electrical performance of our improved films by fabricating monolayer *p*-type $WSe_2$ transistors (**Figures 3a,b**) using an optimized fabrication (described in **Supplementary**

**Figure 6** and **Supplementary Note 2**). Drain current versus gate voltage curves (**Figures 3c-d**) show good enhancement-mode (normally-off) behavior with small, negative threshold voltage, $V_T$. Device-to-device variation in $V_T$ is relatively low, with a standard deviation of $\sigma_{V_t} = 204$ mV for short-channel (50 nm) devices and $\sigma_{V_t} = 108$ mV for long-channel (1 μm) devices. The larger spread in $V_T$ for short-channel devices likely reflects contact resistance variability, while the long-channel devices benefit more directly from the uniformity and quality of the WSe$_2$ channel (contacts play a lesser effect on their behavior), reaching high on-currents above 100 μA μm$^{-1}$. The reported devices were sampled across the chip, and the low variability is consistent with the uniformity established from structural and optical characterization. Additional electrical data are provided in **Supplementary Figures 7** for back-gated devices on SiO$_2$.

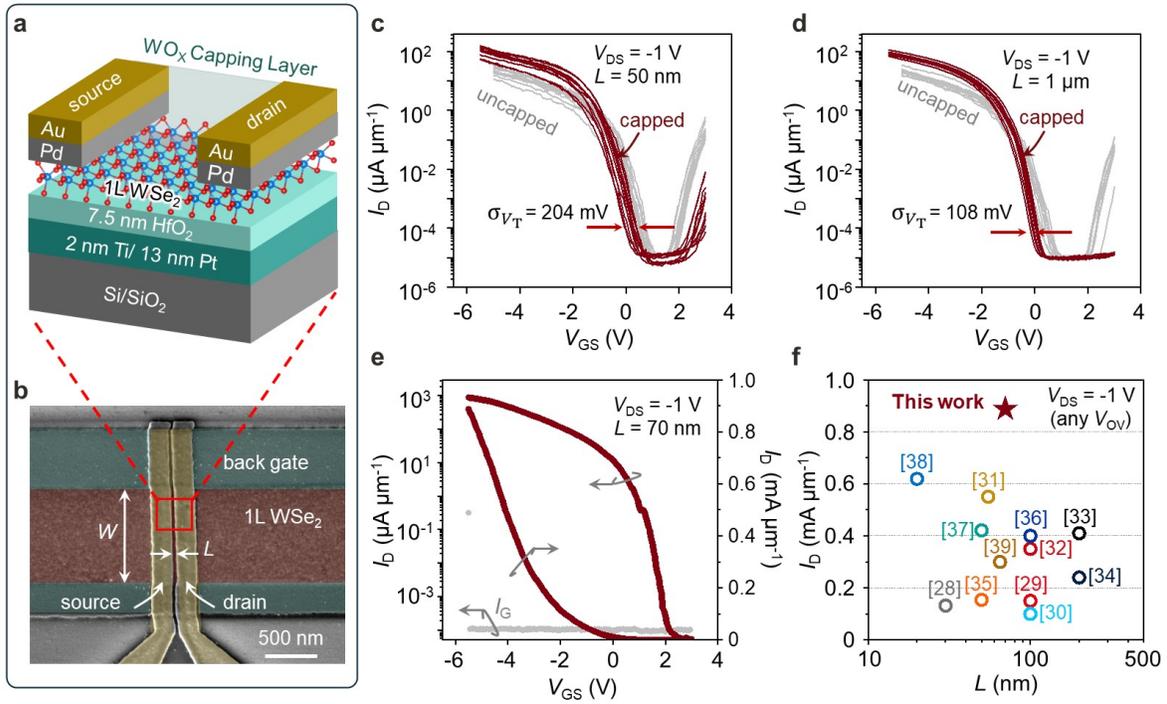

**Figure 3: Monolayer WSe$_2$ transistors. a,** Schematic of our monolayer WSe$_2$ devices. **b,** False-colored, top-down scanning electron microscopy (SEM) image of a representative device, showing the source/drain electrodes in yellow, the WSe$_2$ channel in red, and the local back-gate in green. The width ($W$) measures 950 nm (measured via SEM) and the channel length ($L$) is measured from source to drain. **c,** Measured drain current ($I_D$) vs. gate voltage ($V_{GS}$) of 14 devices with $L \approx 50$ nm short-channel and, **d,** nine devices with $L \approx 1$ μm long-channel. Light grey represents measurements before capping, and dark red represents measurements after capping with WO$_x$. Only the forward sweep is shown for clarity. **e,** Electrical data of 70 nm transistor with highest current density of ~888 μA μm$^{-1}$ at $V_{DS} = -1$ V. This device has a positive-shifted threshold voltage after repeated cycling. **f,** Benchmark of maximum hole current in monolayer WSe$_2$ devices vs. channel length.[28-39] Only devices reaching $I_{max} > 100$ μA μm$^{-1}$ and on/off ratios $>10^3$ are shown. All devices are at $V_{DS} = -1$ V, and at their reported maximum overdrive ($V_{OV} = V_{GS} - V_T$).

To assess the limits of current density, we reduced the Schottky barrier width using an evaporated $WO_x$ capping layer and applied bias-stress to shift $V_T$. Under these conditions, which allow us to reach higher carrier densities, the contact resistance is 404 ± 126 Ω·μm, among the lowest values reported to date for monolayer $WSe_2$ (**Supplementary Figure 8**). In **Figure 3e**, we show a 70 nm channel device that reaches an on-state current of 888 μA μm$^{-1}$ at $V_{DS}$ = -1 V. This represents the highest on-current density reported to date for monolayer $WSe_2$ (**Figure 3f**), comparable to the performance of state-of-the-art monolayer *n*-type $MoS_2$ transistors.[16,33]

## Evaluating the defect density

To understand how this record device performance relates to the underlying material quality, we quantified point defect populations in our monolayer $WSe_2$ films. Scanning transmission electron microscopy (STEM, **Figure 4a**) confirms high crystallinity at the nanometer scale. We measured point defect concentrations using conductive atomic force microscopy (CAFM) and lateral force microscopy (LFM). CAFM maps the picoampere-scale current between the probe and sample, detecting both uncharged (*i.e.,* isovalent) and charged point defects, but requires $WSe_2$ transfer to a conductive substrate.[40] In contrast, LFM detects frictional signatures of individual top-surface defects directly on the as-grown films. Because it only probes the top chalcogen layer, LFM defect counts were doubled to estimate the total density.[41]

High-magnification LFM and CAFM images (**Figures 4b,c**, respectively) yield an isovalent defect density of 5.4 ± 0.4 × 10$^{12}$ cm$^{-2}$ (**Supplementary Figure 9**). CAFM also shows an exceptionally low charged defect density of 5.2 ± 1.1 × 10$^9$ cm$^{-2}$ (**Figure 4d**), comparable to the lowest values reported in self-flux-grown bulk crystals.[6] Achieving such ultralow charged defect concentrations in centimeter-scale films grown by a one-hour CVD process represents a major advance for scalable 2D semiconductors. The commercial $WO_3$ powder was used as-received, suggesting that further defect reduction may be possible through precursor purification and reactor optimization.[10,42]

These quantitative results can be linked to transistor performance.[43,44] **Figure 4e** compares maximum on-current density ($I_{max}$) across monolayer 2D semiconductor studies that also reported defect densities. $I_{max}$ is used here (and in **Figure 3f**) rather than mobility because it includes the effect of contact resistance, avoids extraction artifacts, and directly reflects circuit-level behavior (intrinsic delay scales inversely with $I_{max}$). Although few such datasets exist, the emerging trend suggests that low total defect density alone is insufficient. High performance requires both ultralow charged defect densities (< 10$^{10}$ cm$^{-2}$) and reduced isovalent defects.

At the mechanistic level, contacts and channels are influenced by different defect populations. Atomic-scale studies show Se vacancies are passivated by O upon air exposure, forming isovalent $O_{Se}$ centers[13,50] that shift the valence band edge towards the Fermi level.[51] For *p*-type TMDs, $O_{Se}$ substitutions are predicted to enhance hole injection by lowering the Schottky barrier.[52] However, further reducing $O_{Se}$ density may not improve channel transport, because phonon-limited mobility is expected to plateau at $O_{Se}$ densities below $10^{11}$–$10^{12}$ cm$^{-2}$.[5,53] In contrast, charged defects and impurities such as transition metal substitutions (*e.g.,* Si on W sites) introduce mid-gap states[51] and scatter carriers.[54] Our results therefore suggest that the record-low charged defect density in our films underpins the low channel resistance, while a modest density of $O_{Se}$ substitutions may improve contact injection.[52]

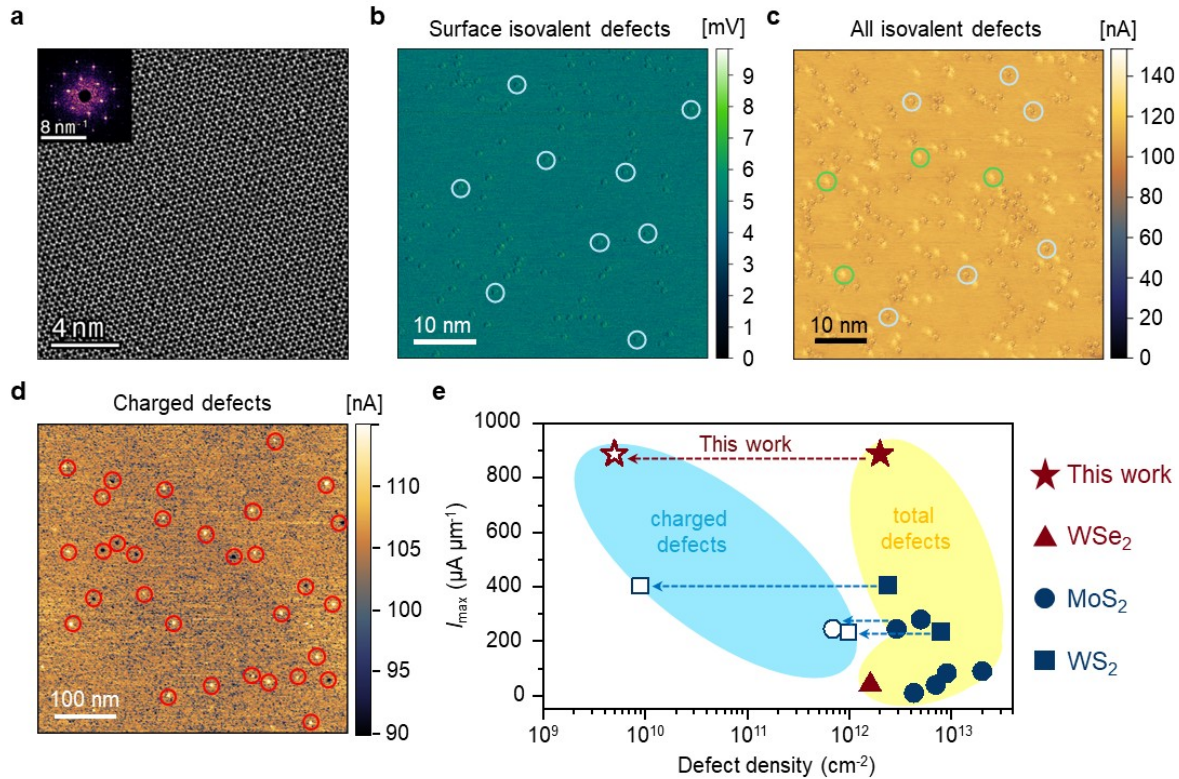

**Figure 4: Defect density measurements. a,** High-resolution scanning transmission electron microscopy (STEM) image of 1L WSe$_2$, demonstrating the crystallinity of the film. Inset shows the fast Fourier transform of the lattice. **b,** Lateral force microscopy (LFM) image of 1L WSe$_2$, where a subset of surface isovalent defects (*i.e.,* at the top side of the WSe$_2$ film) are circled in light blue. **c,** Conductive atomic force microscopy (CAFM) image of the same 1L film, where a subset of top-side isovalent defects is circled in light blue and a subset of bottom-side isovalent defects are circled in light green. **d,** CAFM scan of similar 1L WSe$_2$, with charged defects circled in red. Note the different scale bar in this image. **e,** Plot of transistor on-current density vs. defect density reported in the 2D monolayer film used. All devices have $I_{max}$ measured at $V_{DS} = \pm 1$ V but at any overdrive ($V_{OV}$). Filled symbols correspond to total defect density, open symbols mark charged defect density.[4,10,12,45-49] Red symbols are *p*-type devices, blue are *n*-type.

## Conclusion

The PRC-based precursor engineering strategy enables large-area growth of monolayer WSe$_2$ with an ultralow charged defect density (~5 × 10$^9$ cm$^{-2}$). With optimized fabrication, these films yield *p*-channel transistors that achieve uniform threshold voltages and record on-state hole current density up to 888 μA μm$^{-1}$, closing the longstanding *p*–*n* performance gap in monolayer TMDs. This work establishes a pathway toward integration of low-power complementary 2D semiconductor logic in next-generation electronics.

# Methods

**SSCVD growth.**

Monolayer WSe$_2$ was synthesized in a two-zone CVD system (2-inch horizontal quartz tube) using WO$_3$ (99.995%, Sigma Aldrich, catalog #204781) and Se (99.999%, Thermo Scientific, catalog #010603) as precursors. Se pellets were placed in Zone 1, and the substrate was positioned on a quartz platen in Zone 2, 28 cm downstream from the Se source. Sapphire c-plane substrates were used after sonication in acetone and isopropanol. For the initial growth, WO$_3$ powder was spread on the platen, which seasons the reactor but often yields suboptimal films. Subsequent growths employed the platen recovery cycle (PRC), in which residual material from prior runs was redistributed across the platen, supplemented with fresh WO$_3$, and annealed at 1000 °C under Ar flow. After pump-down to ~10$^{-3}$ Torr, the system was backfilled with Ar to 760 Torr and maintained at this pressure throughout growth. Zone 2 was heated to 880 °C at 44 °C min$^{-1}$, while Zone 1 heating was delayed by 5 min before ramping to 500 °C at 50 °C min$^{-1}$. Growth proceeded for 40 min under 25 sccm Ar and 5 sccm H$_2$, followed by cooling Zone 2 to 460 °C over 30 min and rapid quenching by opening the furnace lid. To prevent H$_2$-induced etching, only Ar was flowed during cooling.

**Material characterization.**

Room-temperature Raman spectroscopy and photoluminescence were performed using a Horiba Labram instrument with a 532 nm laser and 1800 and 600 gr/mm grating, at a laser power of 0.12 mW and spot size of ~0.5 μm.

High-resolution STEM data were acquired at 80kV on the TEAM 0.5, a double-aberration-corrected modified FEI Titan 80-300 microscope at the Molecular Foundry at Lawrence Berkeley National Laboratory. The data were collected on WSe$_2$ samples overlaid on silicon nitride grids. The probe was formed using a 30 mrad convergence angle, and the data were acquired with an annular dark-field detector using an approximate dose of 3x10$^5$ e$^-$/Å$^2$. Images were collected with 0°/90° pairs for post-processing drift correction in the open-source package quantem based on previously published drift correction algorithms.[55,56] After drift correction, the image in Figure 4A was high-pass filtered using an 8$^{th}$ order Butterworth filter at 0.12 Å$^{-1}$ and low-pass filtered using a Gaussian mask with a standard deviation of 5.26 Å$^{-1}$.

Conductive AFM was done using an Asylum Cypher ES using all-in-one diamond tips (spring constant of 6.5 N/m) at a scan rate of 9.77 Hz with a pixel count of 1024 × 1024 for an image size of 50 × 50 nm$^2$ for isovalent defect counting, and at a scan rate of 0.5~1 Hz with a pixel count of 1024 × 1024 for an image size of 1 × 1 μm$^2$ for charged defect counting. Scans are performed until approximately 400 defects are counted, to achieve a ± 10 % error. Lateral force microscopy (LFM) measurements were done on the same instrument, using SCONT tips (spring constant of 0.01 N/m) at a rate of 3.26 Hz with a pixel count of 2048 × 2048 for an image size of 50 × 50 nm$^2$, for isovalent defect counting. All scans were done in ambient, at 25 °C. For CAFM measurements, monolayer samples were transferred onto a graphite/Au/SiO$_2$ substrate using a liquid-assisted PDMS method, where the liquid is a mixture of isopropyl alcohol (IPA) and DI water. Graphite is mechanically exfoliated onto the Au/SiO$_2$ substrate to serve as a conductive ground. The Au/SiO$_2$ substrate is then connected to the sample puck using silver paint to ensure good electrical contact. For LFM measurements, samples are characterized directly on their as-grown substrates.

X-ray photoelectron spectra were collected using a PHI VersaProbe 4 instrument equipped with an Al Kα (1486.6 eV) X-ray source operated at 50 W with a 200 μm spot size. Pass energies of 224 eV (for survey scans) and 55 eV (for core-level scans) were used, providing experimental resolutions of 1.4 eV and 0.8 eV, respectively. A dual-beam charge neutralization system (low-energy ions and electrons) was employed to minimize sample charging. No sputtering was performed during this experiment. Our data set was calibrated using the W 4f W-O component rather than adventitious carbon (commonly referenced at 284.8 eV). Although widely used, calibration to adventitious carbon is well-documented in the literature as a practice that should be avoided.[57,58] All XPS spectra were analyzed using Igor Pro 8.04 and fitted with Voigt functions after Shirley background subtraction.

**Acknowledgements**

A.T.H., A.J.M, and E.P. acknowledge the National Science Foundation (NSF) FuSe2 program (Award 2425218) and support from Applied Materials and TSMC under the Stanford SystemX Alliance. Device fabrication was primarily supported by SUPREME, a JUMP 2.0 Center sponsored by the Semiconductor Research Corporation (SRC) and DARPA. K.N. acknowledges the Stanford Graduate Fellowship and the NSF Graduate Research Fellowship Program. T.P. and E.P. acknowledge support from Intel Corporation. T.P. would like to thank the NSF MPS-Ascend postdoctoral fellowship. A.E.O.P. acknowledges the Knut and Alice Wallenberg Foundation (grant 2022.0374). Fabrication and characterization were primarily performed at the Stanford Nanofabrication Facility (SNF) and the Stanford Nano Shared Facilities (SNSF), supported by the NSF award ECCS-2026822. Work at the Molecular Foundry was supported by the Office of Science, Office of Basic Energy Sciences, of the U.S. Department of Energy under Contract No. DE-AC02-05CH11231. K.X., Y.Y., and M.R.R. acknowledge support from NSF DMR-2340398.

**Author Contributions**

A.T.H and K.N. contributed equally and conceived the project, with guidance from A.J.M. and E.P. A.T.H synthesized WSe$_2$ and characterized material quality using spectroscopic methods. K.N. conducted transistor fabrication and measurements, with deposition and transfer assistance from T.P, and measurements with help from A.E.O.P and Y.S.S. K.X. performed CAFM, and Y.Y. performed LFM scans with M.R.R. A.T.H. and G.D. measured XPS. S.R and W.M performed